\documentclass{iaus}
\usepackage{graphicx}

\title[JD 11.~~Statistical Photometric Parallax] {Determining distances to stars statistically from photometry}

\author[Heidi Jo Newberg]   
{Heidi Jo Newberg$^1$}

\affiliation{$^1$Rensselaer Polytechnic Institute, Dept. of Physics, Applied Physics \& Astronomy, 110 $8^{\rm th}$ St., Troy, NY 12180, USA \\ email: {\tt newbeh@rpi.edu} \\[\affilskip]
}

\pubyear{2012}
\volume{xxx}  
\pagerange{119--126}
\jname{Title of your IAU Symposium}
\editors{A.C. Editor, B.D. Editor \& C.E. Editor, eds.}
\begin{document}

\maketitle

\begin{abstract}
In determining the distances to stars within the Milky Way galaxy, one often uses photometric or spectroscopic parallax.  
In these methods, the type of each individual star is determined, and the absolute magnitude of that star type is compared
with the measured apparent magnitude to determine individual distances.  In this article, we define the term 
{\it statistical photometric parallax}, in which statistical knowledge of the absolute magnitudes of stellar populations
is used to determine the underlying density distributions of those stars.  This technique has been used to determine
the density distribution of the Milky Way stellar halo and its component tidal streams, using very large samples of
stars from the Sloan Digital Sky Survey.  Most recently, the volunteer computing platform MilkyWay@home has been used
to find the best fit model parameters for the density of these halo stars.
\end{abstract}

\firstsection 
\section{Introduction}

\begin{figure}[b]
\begin{center}
 \includegraphics[width=4in]{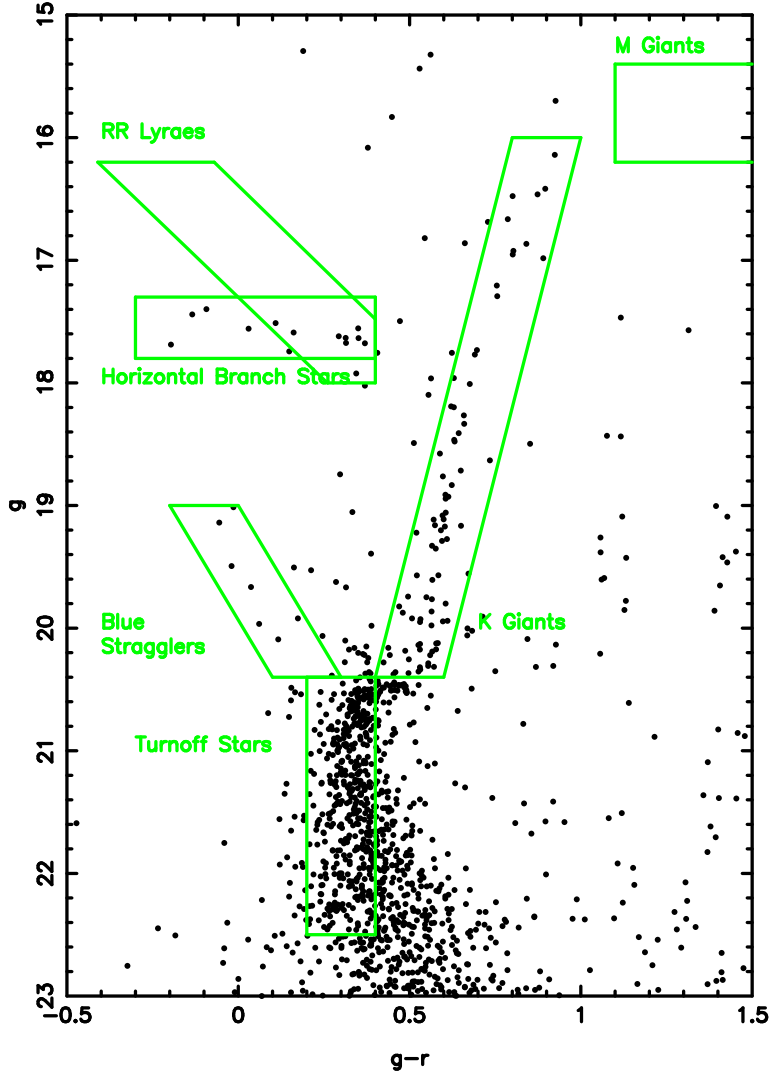} 
 \caption{Color-magnitude diagram of stars withing 4 arcminutes of the globular cluster Palomar 5.  All stars within
four arcminutes of Pal5 were selected from SDSS DR9 (SDSS III Collaboration et al. 2012).  We 
illustrate how sparse the brighter stars are in a stellar population,
compared to the main sequence and turnoff.  Most of the selection boxes for each type of star were drawn to include
the representative stars in Palomar 5.  The RR Lyrae selection box was taken from Rave et al. (2003).  The M giant selection
box was estimated from Yanny et al. (2009) measurements of M giants in the Sagittarius dwarf tidal stream.  There are no
representatives of these types of stars in the Palomar 5 globular cluster.  Although 
giant stars of all types can be observed at larger distances, many
types of giant stars are specific to particular populations.  K giants
are seen in all but the youngest stellar populations, but these stars are more difficult to separate from dwarf K stars
that have the same colors.  Also, K giant stars are more difficult to use as distance indicators because their absolute
magnitude is a strong function of color (temperature), and is also sensitive to metallicity and age.  The drawback to using
turnoff stars as distance indicators is that they vary in absolute magnitude by two magnitudes.  Statistical photometric
parallax allows us to used them effectively to determine the underlying density of stars even though we cannot accurately measure
the distance to each individual star.}
   \label{fig1}
\end{center}
\end{figure}

We often use the stars in the Milky Way to trace its structure.  The brightest stars, which can be
used to trace Galactic structure to the largest distances, include blue horizontal branch stars; O, B
and A main sequence stars; RR Lyraes; Cepheid variables; red clump stars; K giants; and M giants (see Figure~\ref{fig1}).
All of these stellar types are reasonably well calibrated as distance indicators.  K giant and
red clump stars almost always require spectroscopy to accurately distinguish them from the more numerous main
sequence stars of the same color, and
RR Lyrae stars are usually identified from multiple epochs of photometry.  All of these bright stars
are relatively rare in any stellar population, and many of them are only observed in certain populations.  
For example, blue horizontal branch stars and RR Lyrae stars are only found in old populations, M giants
are only found in relatively metal-rich populations, and O, B, and A main sequence stars are only found in
very young populations (though A stars with main sequence gravities are found in some old populations
as blue stragglers).

Deep, large sky area, multicolor surveys with high accuracy calibrations, like the Sloan Digital Sky Survey 
(SDSS; York et al. 2000) enable new methods for studying the structure of the Milky 
Way.  For example, Ivezi{\'c} et al. (2008) derived a formula that allows us to estimate the temperature and 
metallicity of main sequence stars of type G and later from SDSS photometry.  Since these lower mass
main sequence stars do not evolve in the age of the Universe, their absolute magnitudes are independent
of age.  Photometry cannot tell us the surface gravities of these types of stars (Lenz et al. 1998),
but since the vast majority of the red stars in the photometric survey are main sequence stars, this
is not a major obstacle.  Juri{\'c} et al. (2009) used this technique to measure the distances to 48 million
stars, and then used these distances to determine the density distributions of the disks and halo
at 100 pc to 20 kpc from the Sun, over 6500 square degrees of sky.  Though this analysis used
a large number of stars, and resulted in the measurement of stellar density, the Juri{\'c} analysis was
using photometric parallax; they were determining the distance to each star individually using
photometry.

In this article, we introduce the concept of {\it statistical photometric parallax}.  We have used this
technique most successfully to study the structure of the Galaxy using turnoff stars
(Newberg et al. 2002).  These stars are
by definition brighter than main sequence stars such as those used by Juri{\'c} et al., so they can 
trace the structure of the Milky Way 30 kpc or more from the Sun.  However, the turnoff stars in a 
single stellar population can differ in absolute magnitude by two magnitudes (producing a distance 
error of a factor of 2.5).  We do not have a way to determine the distance to a single turnoff star
with reasonable accuracy.  However, it has been shown that the absolute magnitude distribution of turnoff
stars in halo globular clusters are surprisingly similar to each other (Newby et al. 2011), over 
a metallicity range -2.3 dex $<$ [Fe/H] $<$ -1.2 and ages ranging from 9 to 13.5 Gyr.  Recently,
Grabowski, Newby, \& Newberg (2012) showed that this similarity even holds for the globular cluster
Whiting 1, which is thought to be only 6.5 Gyrs old (Carraro, Zinn \& Moni Bidin 2007).  

This striking similarity between the absolute magnitude distributions of turnoff stars was not
expected.  One expects that younger globular clusters would have brighter, bluer turnoff stars.
Also, one expects that more metal-rich clusters will have dimmer, redder turnoff stars.  As it
turns out, older stars in the Milky Way generally have lower metallicity, and the two effects 
cancel each other.  This appears to be an unanticipated consequence of the Milky Way's 
Age-Metallicity Relationship (AMR - Muratov \& Gnedin 2010, Dotter; Sarajedini \& Anderson 2011).  Apparently, 
the absolute magnitude distribution of turnoff stars is similar over the full age and metallicity range of typical
stellar populations in the Milky Way halo.

We will describe here the general techique of statistical photometric parallax, which can be used
to statistically account for the effects of a range of intrinsic brightnesses of the stellar population which
is being used to trace Milky Way density structure, and can also statistically account for the observational
biases in a survey such as the Sloan Digital Sky Survey.  We have implemented this technique as a search
for the density parameters with the highest likelihood of matching the observed data.  Because this parameter
search can be computationally expensive, we have employed supercomputers and a large volunteer computing
platform, MilkyWay@home, that was built to solve this problem.

\section{Using Statistical Photometric Parallax}

We are defining the term {\it statistical photometric parallax} here for the first time, and it is intended to
apply in general for any case where the statistical distribution of absolute magnitudes is used to find the
underlying density distribution of stars.  However, we will describe here as an example the application of this technique
to determine halo substructure using color-selected F turnoff stars, as used by Newberg et al. (2002), 
Cole et al. (2008), and Newby et al. (2012).

In Newberg et al. (2002), stars with colors $0.1<(g-r)_0<0.3$ and $(u-g)_0>0.4$ were selected as turnoff stars.  The
color range was chosen to be bluer than the turnoff of the thick disk, so that halo stars would preferentially be selected.
In this paper, only the simplest form of statistical photometric parallax was employed.  The distance to the 
Sagittarius dwarf tidal stream was determined by assuming the center of the absolute magnitude distribution of turnoff 
stars in the $g$ filter was $M_g=4.2$.  This number was calculated by comparing the apparent magnitude of the turnoff to
the apparent magnitude of RR Lyrae stars in the same stellar population.  In this example, distances to single stars
were not calculated; instead we made a more accurate determination of distance by looking at the distribution of 
apparent magnitudes of a particular set of stars.

In Cole et al. (2008), an algorithm was presented that allowed us to determine not just the distance to the center of
a stream, but the three dimensional density of turnoff stars.  The technique used maximum likelihood to find the model
parameters $\vec{\mathcal{Q}}$ that make the observed star positions $(l_i, b_i, g_i)$ the most likely.  The Likelihood $\mathcal{L}$
is given by the product of the probability density functions (PDFs) evaluated at all of the star positions:
\[\mathcal{L}=\prod{\rm PDF}(l_i,b_i,g_i|\vec{\mathcal{Q}}).\]
The PDF is constructed by the following steps:
\begin{enumerate}
\item For each stellar component, one assumes a parameterized model (for example a double exponential, NFW, Hernquist, etc.)
for the spatial density.
\item This spatial density is transformed to $(l,b,g)$ coordinates, assuming that the absolute magnitude of each of the stars 
is the average for the population.
\item This density is convolved with the absolute magnitude distribution of the tracers, so that we produce the distribution 
that we expect to observe.
\item This expected distribution is multiplied by the completeness for observing stars of a given apparent magnitude
in a given survey, as a function of apparent magnitude.
\item The resulting distribution is normalized so that the integrated probability of finding a star in the entire volume 
observed is one.
\item The final PDF is the sum of the fraction of stars in each component times the normalized distribution, summed over 
the number of components in the model.  The fraction of stars in each component are also parameters that are fit in the 
maximum likelihood optimization.
\end{enumerate}
Of these steps, the most time-consuming is the calculation of the integral over the volume.  Contrary to first impressions,
the time to calculate the likelihood depends more heavily on the number of sub-volumes into which the survey space needs 
to be divided to achieve an accurate result, than on the number of stars in the dataset.

One then uses an optimization technique to find the model parameters that produce the highest likelihood.  When using a
supercomputer, we usually use conjugate gradient descent.  This algorithm is sequential; one evaluates the likelihood and
the derivatives with respect to each parameter, chooses a direction, then decides how far in that direction to go before
repeating that process.

\begin{figure}[b]
\begin{center}
 \includegraphics[width=4in, angle=-90]{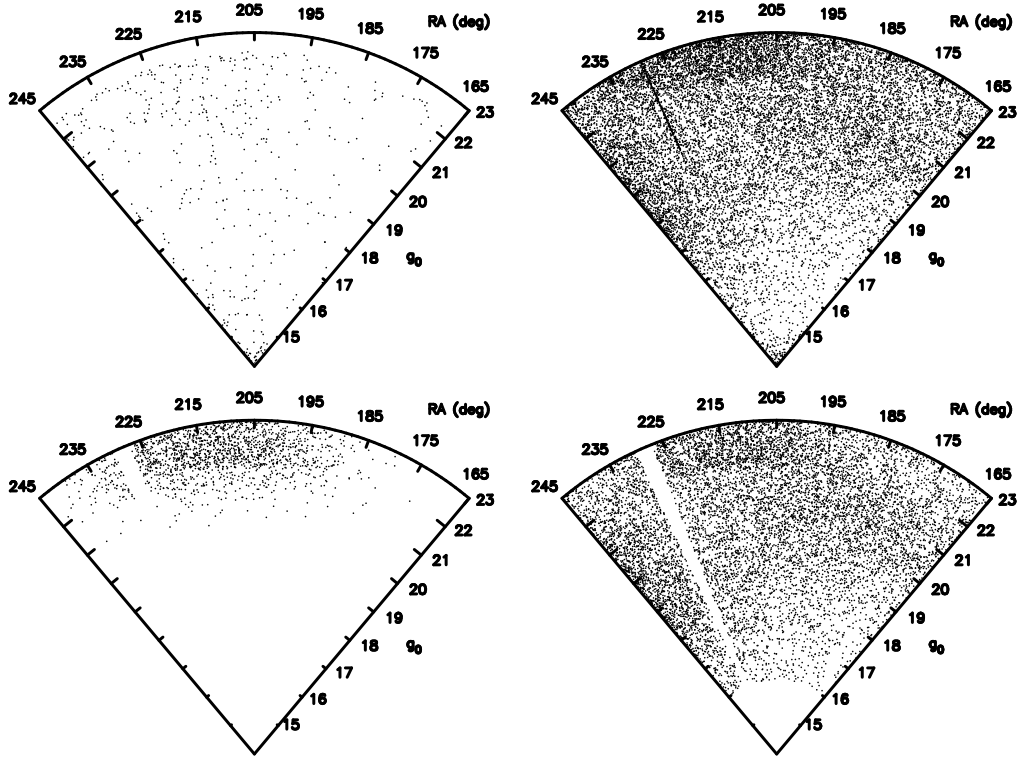} 
\caption{The A and F star distribution on the Celestial Equator.  All of the stars in this figure were selected within
0.1 degrees of the Celestial Equator, and $145^\circ<RA<265^\circ$.  The top left panel shows photometrically 
selected A stars from SDSS DR9 (SDSS-III Collaboration et al. 2012).  The stars were selected with $-0.3<(g-r)_0<0.0$,
$0.8<(u-g)_0<1.5$, and $14<g_0<22.5$, where the subscript
indicates that the values have been corrected for reddening using the extinction calculated in the database.
The top right panel shows photometrically selected turnoff stars, selected from SDSS DR9.  The
stars were selected with $0.2<(g-r)_0<0.4$ and $14<g_0<23$.  Notice that even though we selected only the bluest
turnoff stars, there are very many more blue turnoff stars than stars with the colors of A stars.  However, the wide range
of $g_0$ magnitudes observed in the globular cluster Palomar 5 at $RA=229^\circ$ indicates that the range of absolute
magnitudes of turnoff stars is at least two magnitudes.  Even though the range of absolute magnitudes is large, we can
see the Palomar 5 globular cluster, the Sagittarius dwarf galaxy tidal stream at $(RA,g_0)=(210^\circ,22.5)$, the 
Virgo Overdensity at 
$(RA,g_0)=(180^\circ, 21)$, and stars that are presumably part of the smooth spheroid near the Galactic center at
$(RA,g_0)=(245^\circ, 19)$.  The substructure in A-type stars is not as evident partially because there are many fewer
A-type stars, partially because we have included both blue stragglers and blue horizontal branch stars, and partially because
not all stellar populations have A-type stars.  Blue horizontal branch stars
are about 1.5 magnitudes brighter than turnoff stars, and blue stragglers are about 3.5 magnitudes brigher than
turnoff stars.  
A small number of blue stragglers from the Sagittarius dwarf tidal stream are evident at
$g_0=21$ in the A star panel, but Palomar 5 blue stragglers are not evident.  We do not see blue horizontal branch stars 
from either Palomar 5 or the Sagittarius dwarf tidal stream in this plot.
One can see the advantage of using turnoff stars in identifying the
Sagittarius dwarf tidal stream.  The lower two panels show the stars selected by Newby et al. (2012), from this small
region of the sky, that have the density distribution of the Sagittarius dwarf tidal stream (left) and
the density distribution over everything else (right).  The region around the globular cluster
Palomar 5 was removed from the data before fitting, so that this globular cluster would not affect the results.
}
   \label{fig2}
\end{center}
\end{figure}

Newby et al. (2012) applied this technique to all of the data available in SDSS DR7 to find the density of the Sagittarius
dwarf tidal stream in the North Galactic Cap, and in three SDSS stripes in the South Galactic Cap.  One of the advantages
of this probablistic technique is that we are able to extract from the sample of 1.7 million turnoff stars a set of $200,000$
stars that have the spatial characteristics of the Sagittarius dwarf tidal stream.  This is accomplished by 
generating a random mumber for each star, and using that random number to place it either in the Sagittarius dwarf 
tidal stream catalog, with the probability that a star at that position in the Galaxy is in the Sagittarius dwarf tidal stream;
or in the catalog of non-Sagittarius halo stars, with the probability that a star at that position in the Galaxy is
not in the Sagittarius dwaf tidal stream.  Note that if you wanted to find actual stars in the
Sagittarius dwarf tidal stream, say for spectroscopic follow-up, you should use the original catalog of stars and select
those with the highest probability of being in the Sagittarius dwarf tidal stream.  However, the statistically
separated catalog we generated facilitates the study of density substructures in the halo.  In particular, we can remove
the Sagittarius stream from the original stellar sample so we can study the smaller tidal streams, and the density
structure of the smooth component of the halo.  An example stripe analyzed in Newby et al. (2012) is shown in Figure~\ref{fig2}.

In Cole et al. (2008) and Newby et al. (2012) we modeled the distribution of turnoff star absolute magnitudes as a 
Gaussian centered at $M_g=4.2$ with a width of $\sigma=0.6$ magnitudes.  Since then, we have recognized that the distribution
is asymmetric; there are more stars fainter than the maximum than there are brighter than the maximum.  More importantly we have
learned that, due to larger color errors at fainter magnitudes, the absolute magnitude distribution of color-selected
stars is different near the survey limit than it is for the brighter stars.  This effect is much larger than we expected it
to be.  Near the survey limit, the majority of the stars are not turnoff stars, but are fainter main sequence stars that
have scattered into our color selection limits due to large measurement errors.  Because we are using a statistical
approach, this effect can be included in the analysis by varying the absolute magnitude distribution as a function of
apparent magnitude.  We plan to include this in future analyses.

\section{Processing time and the MilkyWay@home Volunteer Computing Platform}

We originally tried to implement the search for maximumum likelihood on a single CPU.  In a single $2.5^\circ$-wide SDSS
data stripe, we fit 2 parameters to a smooth halo with a Hernquist profile, and 6 parameters per tidal debris stream.
If there were three tidal streams in a single stripe, there would be 20 parameters.  Evaluating the likelihood for one
guess for the model parameters currently takes about 4 hours, with most of the time spent integrating the PDF over the survey volume.
In order to optimize 20 parameters requires about 50 likelihood evaluations per conjugate gradient descent step and
50 steps per maximum likelihood evaluation.  This totals ten thousand hours per optimization (over a year).

Luckily the optimization is embarrassingly parallel.  It is possible to parallelize the integral, since each integral volume
calculation is completely independent of the others.  We are able to run this algorithm on a 256 node rack of
a Blue Gene/L supercomputer.  Parallelizing the integral over 256 nodes cuts the time per likelihood calculation down under 
a minute.  A conjugate gradient step can then be accomplished in
47 minutes, and ten iterations can be accomplished in under eight hours (which is comfortably less the the maximum job
size allowed in our queue).  In practice, we need to try several conjugate gradient descents to approximate the best
parameters.  Once they are known approximately, we run of order ten conjugate gradient descents starting near the best
values.  Including submitting jobs a few at a time and waiting for queue time, this process can take a couple of weeks to 
validate the results for one SDSS stripe.

Currently our best method for computing the parameters is the volunteer computing platform MilkyWay@home.  This project is
part of the Berkeley Open Infrastructure for Network Computing (BOINC; Anderson, Korpela \& Walton 2005) group of volunteer 
computing platforms.  The first
and most famous of these is SETI@home.  BOINC offers us a template server and database application, and an infrastructure
for volunteers to donate their time to our server.  We implemented our own server, including the maximum likelihood
algorithm, a set of optimization routines that will run in a heterogeneous, asynchronous parallel computing environment
(Desell et al. 2010b),
and a modified server application that sends out ``work units" to the volunteers, collects the results, and validates that
the results are not in error (Desell et al. 2010a).  One of the surprises in operating a BOINC server is that 
some of the results sent back from 
the volunteers are not correct, either because their hardware is malfunctioning, they did not update the software
correctly, or they purposely sent back a wrong answer quickly so that they can accumulate BOINC ``credit" more quickly.
It is impossible to overestimate how important it is to our volunteers that they get credit for the work units their
computers crunch, and that credit is apportioned fairly between the volunteers.

MilkyWay@home is currently delivering 0.5 PetaFLOPS of computing power from 25,000 active volunteers giving us access to
over 35,000 CPUs or GPUs.  The majority of the computing power comes from the GPUs, the best of which can process our likelihood
calculations about 100 times faster than the CPUs (Desell et al. 2009).  It is not easy to parallelize the computation 
of the integral on BOINC,
because that would require communication between the processors, which is not possible at this time.  Instead, we parallelize
the calculations by sending a single likelihood calculation (including the whole integral for a given 
$\vec{\mathcal{Q}}$) to each volunteer.  These work
units can take a couple minutes (if the work unit is sent to a GPU) or four or more hours (if it goes to a CPU), but it might take
minutes, days, or weeks for the likelihood to be returned depending on how much the volunteer is using the computer for her
own purposes, and whether she turns it off.  

It takes far more computing power to calculate best fit parameters
on the MilkyWay@home computing system than on a supercomputer.  There are four factors responsible for this: (1) We cannot
use sequential searches like conjugate gradient descent.  Instead, we use ``particle swarm" or a genetic search algorithm.
In the particle swarm technique, we send out a random set of guesses that span the parameter space.  As the likelihood
results from the volunteers come back, we send out more work units with guesses that are closer to the higher likelihoods.
This search method requires many more steps, but produces more accurate results.  (2) A fraction (about 10\%) of the 
work units, selected at random, are 
sent out five times so that we can validate that they are correct.  If three or more are returned with the same answer, then 
the result is validated.  We also choose to validate the best likelihoods (since those influence our future guesses),
and the likelihoods of users that have submitted previous results that did not validate correctly.  (3) Some of the 
work units that are sent out are never returned.  (4) Because we can, we run the searches
for a longer period of time over a wider range of of parameter space and we get better global values for the parameters
(Newby et al. 2012).  To optimize one stripe takes 1-2 weeks, and hundreds of thousands of likelihood calculations.
There are enough volunteers that we can optimize 4-5 stripes at the same time and still get the results within the same
time scale.  By putting more jobs on MilkyWay@home at the same time, we decrease the number of work units that are sent
out at the same time.  By waiting a little longer to send out more work units, the results of previous searches can be used
to make better guesses of the parameters, so adding more jobs increases the computing time at a rate that is less than
linear in the number of jobs.

\section{Discussion and Conclusion}

The purpose of this conference contribution is to define the term {\it statistical photometric parallax}, which allows
us to determine the density of a population of stars, even if we cannot determine the distance to each individual star
in the population.  We find the most likely parameters for the density distribution, given that we know the absolute
magnitude distribution and the observational constraints of the observed sample of stars.  If the population of stars
is all at the same distance (for example in a globular cluster or tidal stream), then statistical photometric
parallax can be used to determine the distance to the stellar population.  Because the SDSS made available a large,
well calibrated sample of stars with multi-color photometry, this technique has recently become feasible.

The example I present is the use of turnoff stars to determine the density distribution of stars in the stellar halo
of the Milky Way.  By apparent coincidence, the distribution of absolute magnitudes of turnoff stars is very similar
for all stellar populations in the age and metallicity range of halo stars.  This appears to be a result of the Milky Way
age-metallicity relationship.  We are in the process of using this technique to accurately map the density distribution
of the entire Milky Way stellar halo.

The drawback to this techique is that to use it one must use a maximum likelihood algorithm that can in some cases 
require high performance parallel computing to get an accurate measurement of the parameters in the density function.  In 
the process of learning to use this method, we created a large volunteer computing platform called MilkyWay@home.

\acknowledgement I would like to thank Matthew Newby and Brian Yanny for helping me obtain the data for the figures 
in this publication.  I also thank Matthew Newby and Jeff Carlin for their help in proofreading.
This paper is based upon work supported by the National Science Foundation under Grant No. AST 10-09670.  I also
would like to thank the MilkyWay@home volunteers for providing us with computing power at no cost, and the Marvin Clan
for their support.
Funding for SDSS-III has been provided by the Alfred P. Sloan Foundation, the Participating Institutions, the National 
Science Foundation, and the U.S. Department of Energy Office of Science. The SDSS-III web site is http://www.sdss3.org/.

\end{document}